\documentclass[journal]{IEEEtran}
\newcommand{\ee}{\epsilon}			
\usepackage{epsf,psfrag,amssymb,amsfonts,color,cite,fancybox}
\usepackage[mathscr]{eucal}
\usepackage[dvips]{graphicx}
\usepackage{ifpdf}
\usepackage{longtable}	
\usepackage{subfigure}
\usepackage{multicol}
\ifCLASSINFOpdf
\else
\fi
\usepackage[cmex10]{amsmath}
\begin{document}
\title{Application of Pseudo-Transient Continuation Method in Dynamic Stability Analysis}

\author{Xiaozhe Wang,  ~\IEEEmembership{Student Member,~IEEE,}
        Hsiao-Dong Chiang,~\IEEEmembership{Fellow,~IEEE.}
        % <-this % stops a space
\thanks{Xiaozhe Wang is with the Department
of Electrical and Computer Engineering, Cornell University, Ithaca,
NY, 14853 USA e-mail: xw264@cornell.edu}% <-this % stops a space
\thanks{Hsiao-Dong Chiang is with the Department of Electrical and Computer Engineering, Cornell University, Ithaca, NY 14853 USA email:hc63@cornell.edu}% <-this % stops a space
%\thanks{Manuscript received ; revised .}
}

\maketitle

\begin{abstract}
\boldmath
In this paper, pseudo-transient continuation method has been modified and implemented in power system long-term stability analysis. This method is a middle ground between integration and steady state calculation, thus is a good compromise between accuracy and efficiency. Pseudo-transient continuation method can be applied in the long-term stability model directly to accelerate simulation speed and can also be implemented in the QSS model to overcome numerical difficulties. Numerical examples show that pseudo-transient continuation method can provide correct approximations for the long-term stability model in terms of trajectories and stability assessment.
\end{abstract}


% \fi
\IEEEpeerreviewmaketitle

\section{introduction}

\IEEEPARstart{T}IME-domain simulation is an important approach for power system dynamic analysis. However, the complete system model, or interchangeably the long-term stability model, typically includes different components where each component requires several differential and algebraic equations (DAE) to represent, at the same time, these dynamics involve different time scales from millisecond to minute. As a result, the total number of DAE of a real power system can be formidably large and complex such that time domain simulation over long time intervals is expensive\cite{Chiang:book}. These constraints are even more stringent in the context of on-line stability assessment.
%generators, exciters, governors, loads, over excitation limiter, load tap changers, transformers and other power electronic devices, in terms of computational efforts and data processing

Intense efforts have been made to accelerate the simulation of long-term stability model. One approach is to use a larger time step size to filter out the fast dynamics or use automatic adjustment of step size according to system behavior in time-domain simulation \cite{Kundur:book}\cite{Kurita:article}\cite{Cutsem:book} from the aspect of numerical method. Another approach is to implement the Quasi Steady-State (QSS) model in long-term stability analysis \cite{Cutsem:book}\cite{Cutsem:artical} from the aspect of model approximation. Nevertheless, the QSS model suffers from numerical difficulties when the model gets close to singularities which were addressed in \cite{Cutsem:artical2}-\cite{Wang:artical}. %On the other hand, the QSS model requires more robust numerical method as more inconsistent initial conditions exist whenever discrete variables jump. %And singularity of the QSS model doesn't necessarily imply the instability of the long-term stability model.
Moreover, the QSS model can not provide correct approximations of the long-term stability model consistently as numerical examples shown in \cite{Wangxz:article}\cite{Wangxz:wiley}. In addition, sufficient conditions of the QSS model were developed in \cite{Wangxz:journal} which pointed to a direction to improve the QSS model. As a result, the QSS model requires improvements in both model development and numerical implementation. This paper contributes to the latter one.
%under which the QSS model can provide correct approximations in terms of trajectories and $\omega$-limit set

%it is well recognized that the QSS model has issues in representing dynamic behaviors of the long-term stability model and can provide incorrect approximations for stability analysis \cite{Cutsem:artical2}\cite{Cutsem:artical3}\cite{Wangxz:article}\cite{Wangxz:journal}. Some counter examples in which the QSS model were stable while the long-term stability model underwent long-term instabilities were presented in \cite{Wangxz:article}\cite{Wangxz:journal}. In addition, a theoretical foundation for the QSS model was presented in \cite{Wangxz:journal} where sufficient conditions of the QSS model for accurate approximation of the long-term stability model in terms of trajectories and $\omega$-limit set were derived.

In this paper, we apply pseudo-transient continuation ($\Psi tc$) which is a theoretical-based numerical method in power system long-term stability analysis. Pseudo-transient continuation method can be implemented directly in the long-term stability model to accelerate simulation speed compared with conventional implicit integration method. On the other hand, the method can also be applied in the QSS model to overcome possible numerical difficulties due to good stability property.

This paper is organized as follows. Section \ref{sectiondyPTC} briefly reviews general pseudo-transient continuation method in DAE system. Section \ref{sectionPTCinpowersystem} includes a introduction about power system models followed by implementation of pseudo-transient continuation method in the long-term stability model and the QSS model respectively. Section \ref{sectionnumerical} presents three numerical examples to show the feasibility of the method. And conclusions are stated in Section \ref{sectionconclusion}. %Section \ref{sectiondymodel} and Section \ref{sectionqssmodel} briefly introduce the basic concept of complete dynamic model and the QSS model with numerical examples. Section \ref{counterexample} presents two counter examples in which the QSS model fails to provide correct approximations of the complete model. Specifically, while the QSS model is stable, the complete model suffers from voltage instabilities. Also, theoretical explanation for this failure is presented. Conclusions and perspectives are stated in Section \ref{conclusion}.
%In both cases, a much faster simulation speed can be achieved compared with traditional implicit integration method.

%we presents a numerical technique--- for time domain simulation which can be implemented for both the long-term stability model and the QSS model. Pseudo-transient continuation method employs adaptive time-stepping to integrate DAEs until

%A set of differential and algebraic equations (DAE) are numerically solved to study the dynamic behavior of power systems. Power systems networks typically include thousands of generators, exciters, governors, loads, transformers, and other power electronics devices, where each individual component may need several differential and algebraic equations to represent; thus, the total number of DAEs of a real power system can be formidably large.

\section{pseudo-transient continuation method}\label{sectiondyPTC}
Pseudo-transient continuation is a physically-motivated method and can be used in temporal integration. % in which the time step is increased as the iteration progresses in order to achieve fast convergence near a solution. can help reach the steady state solution much faster compared with the conventional step-by-step integration
The method follows the solution of dynamical system accurately in early stages until the steady state is approaching. The time step is thereafter increased by sacrificing temporal accuracy to gain rapid convergence to steady state \cite{Kelley:article}. If only the steady state of a dynamical system instead of intermediate trajectories is of interest, pseudo-transient continuation method is a better choice than accurate step-by-step integration. On the other hand, compared with methods that solve nonlinear equations for steady state such as line-search and trust region methods, pseudo-transient continuation method can avoid converging to nonphysical solutions or stagnating when the Jacobian matrix is singular. This is particularly the case when the system has complex features such as discontinuities which exist in power system models.

Therefore, $\Psi tc$ method can be regarded as a middle ground between integrating accurately and calculating the steady state directly. $\Psi tc$ method can help reach the steady state quickly while maintain good accuracy for the intermittent trajectories. For ODE dynamics, sufficient conditions for convergence of $\Psi tc$ were given in \cite{Kelley:article}. The results were further extended the semi-explicit index-one DAE system in \cite{Kelley:article2}. We recall the basic algorithm here.

%Pseudo-transient continuation ($\Psi tc$) is a method for globalizing the computation of steady-state solutions of nonlinear differential equations. In addition, the method employs adaptive time steps to integrate an initial value problem until sufficient accuracy is achieved to switch over to Newton's method and gain a rapid asymptotic convergence.

We consider the following semi-explicit index-one DAE system:
\begin{equation}\label{DAE}
D\left(\begin{array}{c}\dot{u}\\\dot{v}\end{array}\right)=-\left(\begin{array}{c}f(u,v)\\g(u,v)\end{array}\right)=-F(x)
\end{equation}
with initial value $x(0)=x_0$. Here $u\in \Re^{N_1}$, $v\in \Re^{N_2}$, $x=[u,v]^T \in \Re^{N_1+N_2}$, and
\begin{equation}
D=\left(\begin{array}{cc}D_{11}&0\\0&0\end{array}\right)\nonumber
\end{equation}
where $D_{11}$ is a nonsingular scaling matrix. We assume the initial condition for (\ref{DAE}) is consistent, i.e. $g(u(0),v(0))=0$ and seek to find the equilibrium point $x^\star$ such that $F(x^\star)=0$ and satisfies
$\lim_{t\to \infty}x(t)=x^\star$.

As stated before, the step-by-step integration is too time consuming if the intermediate states are not of interest. On the other hand, Newton's method for $F(x)=0$ alone usually fails as the initial condition is not sufficiently near the equilibrium point. %Standard globalization such as line search or trust region methods often stagnate at local minimum of $\|F\|$. If the initial condition is too far away from $x^\star$, the conventional method may stagnate at a singularity of

The $\Psi tc$ procedure is defined by the iteration:
\begin{equation}\label{PTC}
x_{n+1}=x_n-(\delta_n^{-1}D+F'(x_n))^{-1}F(x_n)
\end{equation}
where ${\delta_n}$ is adjusted to efficiently find $x^\star$ rather than to enforce temporal accuracy. The convergence results in \cite{Kelley:article}\cite{Kelley:article2} assume that the time step is updated with "switched evolution relaxation"(SER):
\begin{equation}\label{delta}
\delta_n=\mbox{max}(\delta_{n-1}\frac{\|F(x_{n-1})\|}{\|F(x_{n})\|},\delta_{max})
\end{equation}

The algorithm is shown as below:
\begin{IEEEdescription}[\IEEEusemathlabelsep\IEEEsetlabelwidth{A.}]
\item[\textbf{Algorithm 1 ($\Psi tc$ for general DAE)}]
\item[\textbf{\textit{1.}}] Set $x=x_0$ and $\delta=\delta_0$. Evaluate $F(x)$.
\item[\textbf{\textit{2.}}] While $\|F(x)\|$ is too large.
\begin{LaTeXdescription}
\item[\textbf{\textit{a}}] Solve $(\delta^{-1}D+F'(x))s=-F(x)$.
\item[\textbf{\textit{b}}] Set $x=x+s$.
\item[\textbf{\textit{c}}] Evaluate $F(x)$.
\item[\textbf{\textit{d}}] Update $\delta$ according to (\ref{delta}).
\end{LaTeXdescription}
\end{IEEEdescription}

Step 2.a is a Newton step which is typically solved by an iterative method which terminates on small linear residuals while it may also be solved by inexact Newton iteration. Note that the failure of $\Psi tc$ usually can be well signaled by reaching the bound on the total number of iterations \cite{Kelley:article}.

The convergence of $\Psi tc$ for smooth $F$ was proved in \cite{Kelley:article2} under the assumptions that the DAE has index one in a certain uniform sense, that it has a global solution in time, and that the solution converges to a steady state. The result were further extended to nonsmooth $F$ in \cite{Kelley:article3} with $F'(x_n)$ in (\ref{PTC}) replaced by a generalized derivative. %The $\Psi tc$ iteration remains in a neighborhood of the solution of (\ref{DAE})

Next we explain why $\Psi tc$ has a better stability property. Firstly, conventional integration methods insist on a small norm of the linear residual at each step and will either converge, diverge to infinity, or stagnate at a point where iteration matrix is singular. However, $\Psi tc$ method will accept an increase in the residual, responding to that increase by decreasing $\delta$\cite{Kelley:article2}. In addition, $\Psi tc$ stems from backward Euler method which is an attractive choice when stability is the desired property instead of accuracy.

%Next we explain why the QSS model is more likely to meet numerical problems and how $\Psi tc$ can ease numerical problems for the QSS model. Firstly, the QSS model more tends to meet numerical difficulties because whenever discrete variable $z_d$ jump, there are more algebraic equations whose initial conditions are inconsistent. Thus the QSS model requires a more stable numerical method compared with the long-term stability model. On the other hand, $\Psi tc$ has a better stability property compared to other integration methods. Because conventional integration methods insist on a small norm of the linear residual at each step and will either converge, diverge to infinity, or stagnate at a point where iteration matrix is singular. However, $\Psi tc$ method will accept an increase in the residual, responding to that increase by decreasing $\delta$\cite{Kelley:article2}. In addition, $\Psi tc$ stems from backward Euler method which is an attractive choice when stability is the desired property instead of accuracy.

%especially when the short-term dynamics $x$ are slightly excited as long-term dynamics evolve. Mathematically speaking, if $\|f(z_c,z_d,x,y)\|$ is not small enough, thus $x$ are far from their equilibrium point as $z_c$, $z_d$ moves, then divergence of numerical method may happen.
%When stability is the desired property instead of accuracy, backward Euler will be an attractive choice.

One may think of $\Psi tc$ method as a predictor-corrector method where the simple predictor is from previous time step and the corrector is backward Euler with Newton iteration \cite{Kelley:article}. To see this, consider the implicit Euler step from $x_n$ with $\delta_n$,
\begin{equation}
z_{n+1}=x_n-\delta_n D^{-1}F(z_{n+1})
\end{equation}
Thus, $z_{n+1}$ is the root of the following equation:
\begin{equation}
G(\eta)=\eta+\delta_nD^{-1}F(\eta)-x_n\nonumber
\end{equation}
The Newton's method to find the root of the above equation is:
\begin{equation}
\eta_{k+1}=\eta_k-(I+\delta_n D^{-1}F^\prime(\eta_k))^{-1}(\eta_k+\delta_n D^{-1}F(\eta_k)-x_n)\nonumber
\end{equation}
If we take $\eta_0=x_n$, then by Binomial Inverse Theorem, the first Newton iterate is:
\begin{eqnarray}
\eta_1&=&x_n-(I+\delta_n V^{-1}F^\prime(x_n))^{-1}\delta_nV^{-1}F(x_n)\nonumber\\
&=&x_n-(\delta_n^{-1}V+F^\prime(x_n))^{-1}F(x_n)\nonumber
\end{eqnarray}
which is exactly $\Psi tc$ step. As $\Psi tc$ method has a better stability property, it can be applied in the QSS model when conventional integration method fails to converge.

%On the other hand, the QSS model is more likely to meet numerical difficulties because whenever discrete variable $z_d$ jump, there are more algebraic equations whose initial conditions are inconsistent. As a result, $\Psi tc$ method can be applied in the QSS model when conventional integration method fails to converge. %As the backward Euler method has a better stability compared to other implicit integration method, $\Psi tc$ method can be used to overcome numerical difficulties in the QSS model.
\section{$\Psi tc$ applied in power system models}\label{sectionPTCinpowersystem}
In this section, we firstly introduce power system models in long-term stability analysis. Then we apply $\Psi tc$ method to the long-term stability model with modifications. %such that initial conditions of $\Psi tc$ method are consistent.
Finally, we present an algorithm of $\Psi tc$ method in the QSS model to overcome possible numerical difficulties.
\subsection{Power System Models}
The long-term stability model for calculating system dynamic response relative to a disturbance can be described as:
\begin{eqnarray}
\dot{z}_{c}&=&\ee{h}_c({z_c,z_d,x,y})\label{slow ode}\\
{z}_d(k)&=&{h}_d({z_c,z_d(k-1),x,y})\label{slow dde}\\
\dot{{x}}&=&{f}({z_c,z_d,x,y})\label{fast ode}\\
{0}&=&{g}({z_c,z_d,x,y})\label{algebraic eqn}
\end{eqnarray}

Equation (\ref{algebraic eqn}) describes the transmission system and the internal static behaviors of passive devices, and (\ref{fast ode}) describes the internal dynamics of devices such as generators, their associated control systems, certain loads, and other dynamically modeled components. ${f}$ and ${g}$ are continuous functions, and vector ${x}$ and ${y}$ are the corresponding short-term state variables and algebraic variables. Besides, Equations (\ref{slow ode}) and (\ref{slow dde}) describe long-term dynamics including exponential recovery load, turbine governor, load tap changer (LTC), over excitation limiter (OXL), etc. ${z}_c$ and ${z}_d$ are the continuous and discrete long-term state variables respectively, and $1/\ee$ is the maximum time constant among devices. Note that shunt compensation switching and LTC operation are typical discrete events captured by (\ref{slow dde}) and $z_d$ is shunt susceptance and the transformer ratio correspondingly. Transitions of $z_d$ depend on system variables, thus $z_d$ change values from $z_d(k-1)$ to $z_d(k)$ at distinct times $t_k$ where $k=1,2,3,...N$, otherwise, these variables remain constants. Since short-term dynamics have much smaller time constants compared with those of long-term dynamics, $z_c$ and $z_d$ are also termed as slow state variables, and $x$ are termed as fast state variables.

If we represent the long-term stability model and the QSS model in $\tau$ time scale, where $\tau=t\ee$, and we denote $\prime$ as $\frac{d}{d\tau}$, then the long-term stability model of power system can be represented as:
\begin{eqnarray}\label{complete}
{z}_{c}^\prime&=&{h}_c({z_c,z_d,x,y}),\hspace{0.55in}{z_c(\tau_0)=z_{c0}}\\
z_d(k)&=&h_d(z_c,z_d(k-1),x,y),\quad z_d(\tau_0)=z_{d}(0)\nonumber\\
\ee{x}^\prime&=&{f}({z_c,z_d,x,y}),\hspace{0.65in}{x(\tau_0)=x_0^l}\nonumber\\
{0}&=&{g}({z_c,z_d,x,y})\nonumber%,\qquad\quad {y(\tau_0)=y_0}\nonumber
\end{eqnarray}
where the study region $(z_c,z_d,x,y)\in U={D_{z_c}}\times{D_{z_d}}\times{D_{x}}\times{D_{y}}$, and $D_{z_c}\subseteq\Re^p$, $D_{z_d}\subseteq\Re^q$, $D_x\subseteq\Re^m$, $D_y\subseteq\Re^n$.

At the same time, the QSS model can be represented as:
\begin{eqnarray}\label{QSS}
{z}_{c}^\prime&=&{h}_c({z_c,z_d,x,y}),\hspace{0.55in}{z_c(\tau_0)=z_{c0}}\\
z_d(k)&=&h_d(z_c,z_d(k-1),x,y),\quad z_d(\tau_0)=z_{d}(0)\nonumber\\
{0}&=&{f}({z_c,z_d,x,y})\nonumber\\%,\qquad\quad{x(\tau_0)=x_0}\nonumber\\
{0}&=&{g}({z_c,z_d,x,y})\nonumber%,\qquad\quad {y(\tau_0)=y_0}\nonumber
\end{eqnarray}
%and transient stability models are:
%\begin{eqnarray}
%\dot{x}&=&{f}({z_c,z_d,x,y})\\%,\hspace{0.65in}{x(\tau_0)=x_0^l}\nonumber\\
%{0}&=&{g}({z_c,z_d,x,y})\nonumber%,\qquad\quad {y(\tau_0)=y_0}\nonumber
%\end{eqnarray}
%The transient stability model and the QSS model are regarded as two approximations of the long-term stability model in short-term and long-term time scales respectively, and they are believed to offer a good compromise between accuracy and efficiency. In transient stability model, slow variables are considered as constants. While in the QSS model, the dynamic behavior of fast variables are considered as instantaneously fast and thus replaced by its equilibrium equations in long-term time scale.

%where $\tau=t\ee$.

Moreover, the long-term stability model (\ref{complete}) can be regarded as two decoupled systems (\ref{couple1}) and (\ref{couple2}) showed as below when $z_d$ jump from $z_d(k-1)$ to $z_d(k)$:
\begin{equation}
z_d(k)=h_d(z_c,z_d(k-1),x,y),\qquad z_d(\tau_0)=z_{d}(k-1) \label{couple1}
\end{equation}
and
\begin{eqnarray}\label{couple2}
{z}_{c}^\prime&=&{h}_c({z_c,z_d(k),x,y}),\hspace{0.34in}{z_c(\tau_0)={z}_{ck}}\\
\ee{x}^\prime&=&{f}({z_c,z_d(k),x,y}),\qquad\quad{x(\tau_0)=x_k^l}\nonumber\\
{0}&=&{g}({z_c,z_d(k),x,y})\nonumber%,\qquad\quad {y(\tau_0)=y_0}\nonumber
\end{eqnarray}
discrete variables $z_d$ are updated first and then system (\ref{couple2}) works with fixed parameters $z_d$.

Similarly, when $z_d$ jump from $z_d(k-1)$ to $z_d(k)$, the QSS model (\ref{QSS}) can be decoupled as:
\begin{equation}
z_d(k)=h_d(z_c,z_d(k-1),x,y),\qquad z_d(\tau_0)=z_{d}(k-1) \label{coupleqss1}
\end{equation}
and
\begin{eqnarray}\label{coupleqss2}
{z}_{c}^\prime&=&{h}_c({z_c,z_d(k),x,y}),\qquad\quad {z_c(\tau_0)=z_{ck}}\\
{0}&=&{f}({z_c,z_d(k),x,y})\nonumber\\%,\qquad\quad{x(\tau_0)=x_0}\nonumber\\
{0}&=&{g}({z_c,z_d(k),x,y})\nonumber%,\qquad\quad {y(\tau_0)=y_0}\nonumber
\end{eqnarray}

\subsection{$\Psi tc$ for the long-term stability model}
Assuming $D_y g$ is nonsingular, then the long-term stability model (\ref{couple2}) with $z_d$ fixed as parameters is a semi-explicit index-1 DAE system. $\Psi tc$ method requires initial condition to satisfy the algebraic equations, however, the discrete equation (\ref{couple1}) will violate this condition whenever it works. As a result, we need to modify the original $\Psi tc$ method for its implementation in long-term stability model.

In power system long-term stability model, $F=[-h_c,-f,-g]^T$, $p=[z_c,x,y]^T\in\Re^{p+m+n}$, $D_1=\left(\begin{array}{cc}I&0\\0&0\end{array}\right)$, where $I$ is the identiy matrix of size $p+m$. In order to make the initial condition of $\Psi tc$ consistent, we switch back to implicit integration method whenever discrete variables jump and set the step length to be $\delta_{0}$. Moreover, $\Psi tc$ method is implemented for the post-fault system starting from $t_0$---several seconds after the contingency. In examples of this paper, $t_0$ was set to be $5s$. The proposed algorithm is shown as below.
\begin{IEEEdescription}[\IEEEusemathlabelsep\IEEEsetlabelwidth{A.}]
\item[\textbf{Algorithm 2 ($\Psi tc$ in long-term stability model)}]
\item[\textbf{\textit{1.}}] Run the long-term stability model up to $t_0$ by implicit integration method. Set the value $(z_c,x,y)$ at $t_0$ as the initial condition $p_0$ of $\Psi tc$, and set $\delta=\delta_0$.
\item[\textbf{\textit{2.}}] While $\|F(p)\|$ is too large.
\begin{LaTeXdescription}
\item[\textbf{\textit{a}}] If discrete variables jump at $t_k$
\begin{LaTeXdescription}
\item[\textbf{\textit{ }}]              Update $z_d$ according to (\ref{couple1}).
\item[\textbf{\textit{ }}]              Set $p_0=(z_c(t_k),x(t_k),y(t_k))$, $\delta=\delta_0$.
\item[\textbf{\textit{ }}]              Solve the Newton step $As=-H$.
\item[\textbf{\textit{ }}]              Set $p=p+s$.
\item[\textbf{\textit{ }}]              Evaluate $F(p)$.
\end{LaTeXdescription}
\item[\textbf{\textit{b}}] Otherwise
\begin{LaTeXdescription}
\item[\textbf{\textit{ }}] Set $p_0=(z_c(t),x(t),y(t))$.
\item[\textbf{\textit{ }}] Solve the Newton step $(\delta^{-1}D+F'(p))s=-F(p)$.
\item[\textbf{\textit{ }}] Set $p=p+s$.
\item[\textbf{\textit{ }}] Evaluate $F(p)$.
\item[\textbf{\textit{ }}] Update $\delta$ according to (\ref{delta}).
\end{LaTeXdescription}
\end{LaTeXdescription}
\end{IEEEdescription}

Note that $A$ and $H$ depend on the specific integration method used. For instance, if implicit trapezoidal method is used, then
\begin{equation}
A=\left(\begin{array}{ccc}I-0.5\delta_0D_{z_c} h_c&I-0.5\delta_0D_{x} h_c&-0.5\delta_0D_y h_c\\I-0.5\delta_0D_{z_c} f&I-0.5\delta_0D_x f &-0.5\delta_0D_y f\\D_{z_c} g&D_x g&D_y g\end{array}\right)\nonumber\\
\end{equation}
\begin{equation}
H=\left(\begin{array}{c}z_c-z_{cn}-0.5\delta_0(h_c+h_{cn})\\x-x_{n}-0.5\delta_0(f+f_{n})\\g\end{array}\right)\nonumber
\end{equation}

\subsection{$\Psi tc$ for the QSS model}
Assuming $\left(\begin{array}{cc}D_x f&D_y f\\D_x f&D_y g\end{array}\right)$ is nonsingular, then the QSS model (\ref{coupleqss2}) with $z_d$ fixed as parameters is a semi-explicit index-1 DAE system. $\Psi tc$ method requires initial condition to satisfy the algebraic equations, however, the discrete equation (\ref{coupleqss1}) will violate this condition whenever it works. As a result, we need to modify the original $\Psi tc$ method for its implementation in the QSS model.

In the QSS model, $F=[-h_c,-f,-g]^T$, $p=[z_c,x,y]^T\in\Re^{p+m+n}$, $D_2=\left(\begin{array}{cc}I&0\\0&0\end{array}\right)$, where $I$ is the identity matrix of size $p$. In order to make the initial condition of $\Psi tc$ consistent, we switch back to implicit integration method whenever discrete variables jump and set the step length to be $\delta_{0}$. Besides, the QSS model is implemented at $t_1$---when short-term dynamics settle down after the contingency. Usually, $t_1$ can be set as 30s. The proposed algorithm is shown as below.
\begin{IEEEdescription}[\IEEEusemathlabelsep\IEEEsetlabelwidth{A.}]
\item[\textbf{Algorithm 3 ($\Psi tc$ in QSS model)}]
\item[\textbf{\textit{1.}}] Run the long-term stability model up to $t_1$ by implicit integration method. Set the value $(z_c,x,y)$ at $t_1$ as the initial condition $p_0$ of the QSS model, and set $\delta=\delta_0$. Start to run the QSS model.
\item[\textbf{\textit{2.}}] If the QSS model has a numerical difficulty by using implicit integration method, then go to step 3, otherwise, continue with the QSS model.
\item[\textbf{\textit{3.}}] While $\|F(p)\|$ is too large.
\begin{LaTeXdescription}
\item[\textbf{\textit{a}}] If discrete variables jump at $t_k$
\begin{LaTeXdescription}
\item[\textbf{\textit{ }}]              Update $z_d$ according to (\ref{coupleqss1}).
\item[\textbf{\textit{ }}]              Set $p_0=(z_c(t_k),x(t_k),y(t_k))$, $\delta=\delta_0$.
\item[\textbf{\textit{ }}]              Solve the Newton step $As=-H$.
\item[\textbf{\textit{ }}]              Set $p=p+s$.
\item[\textbf{\textit{ }}]              Evaluate $F(p)$.
\end{LaTeXdescription}
\item[\textbf{\textit{b}}] Otherwise
\begin{LaTeXdescription}
\item[\textbf{\textit{ }}] Set $p_0=(z_c(t),x(t),y(t))$.
\item[\textbf{\textit{ }}] Solve the Newton step $(\delta^{-1}D+F'(p))s=-F(p)$.
\item[\textbf{\textit{ }}] Set $p=p+s$.
\item[\textbf{\textit{ }}] Evaluate $F(p)$.
\item[\textbf{\textit{ }}] Update $\delta$ according to (\ref{delta}).
\end{LaTeXdescription}
\end{LaTeXdescription}
\end{IEEEdescription}

Similarly, $A$ and $H$ depend on the specific integration method used. If implicit trapezoidal method is used, then
\begin{equation}
A=\left(\begin{array}{ccc}I-0.5\delta_0D_{z_c} h_c&-0.5\delta_0D_{x} h_c&-0.5\delta_0D_y h_c\\D_{z_c} f&D_x f &D_y f\\D_{z_c} g&D_x g&D_y g\end{array}\right)\nonumber\\
\end{equation}
\begin{equation}
H=\left(\begin{array}{c}z_c-z_{cn}-0.5\delta_0(h_c+h_{cn})\\f\\g\end{array}\right)\nonumber
\end{equation}

\section{Numerical Illustration}\label{sectionnumerical}
In this section, three examples are to be presented. The first two examples were the same 145-bus system in which the QSS model met numerical difficulties during simulation while the long-term stability model was stable in long-term time scale. Firstly, $\Psi tc$ method was implemented in the long-term stability model and the speed was more than 7 times faster than the trapezoidal method. Secondly, when the QSS model by trapezoidal method met difficulty, $\Psi tc$ method was implemented in the QSS model and provided correct approximations and the speed was still more than 5 times faster than the long-term stability model by trapezoidal method. And in the last example which was a 14-bus system, the long-term stability model was unstable and $\Psi tc$ method successfully captured the instability which was signaled by reaching the bound of maximum iteration in the Newton step. All simulations were done in psat-2.1.6\cite{Milano:article}.

\subsection{Numerical Example I}\label{qss_numerical}
The system was a 145-bus system \cite{Vittal:article}. There were exciters and power system stabilizers for each of Generator 1-20. And there were turbine governors for each of Generator 30-40. Besides, there were three load tap changers at lines between Bus 73-74, Bus 73-81 and Bus 90-92 respectively. The contingency was a line loss between Bus 1-6.% The parameters are given in the appendix.

\begin{figure}[!ht]
\centering
\begin{minipage}[t]{0.5\linewidth}
\includegraphics[width=1.8in ,keepaspectratio=true,angle=0]{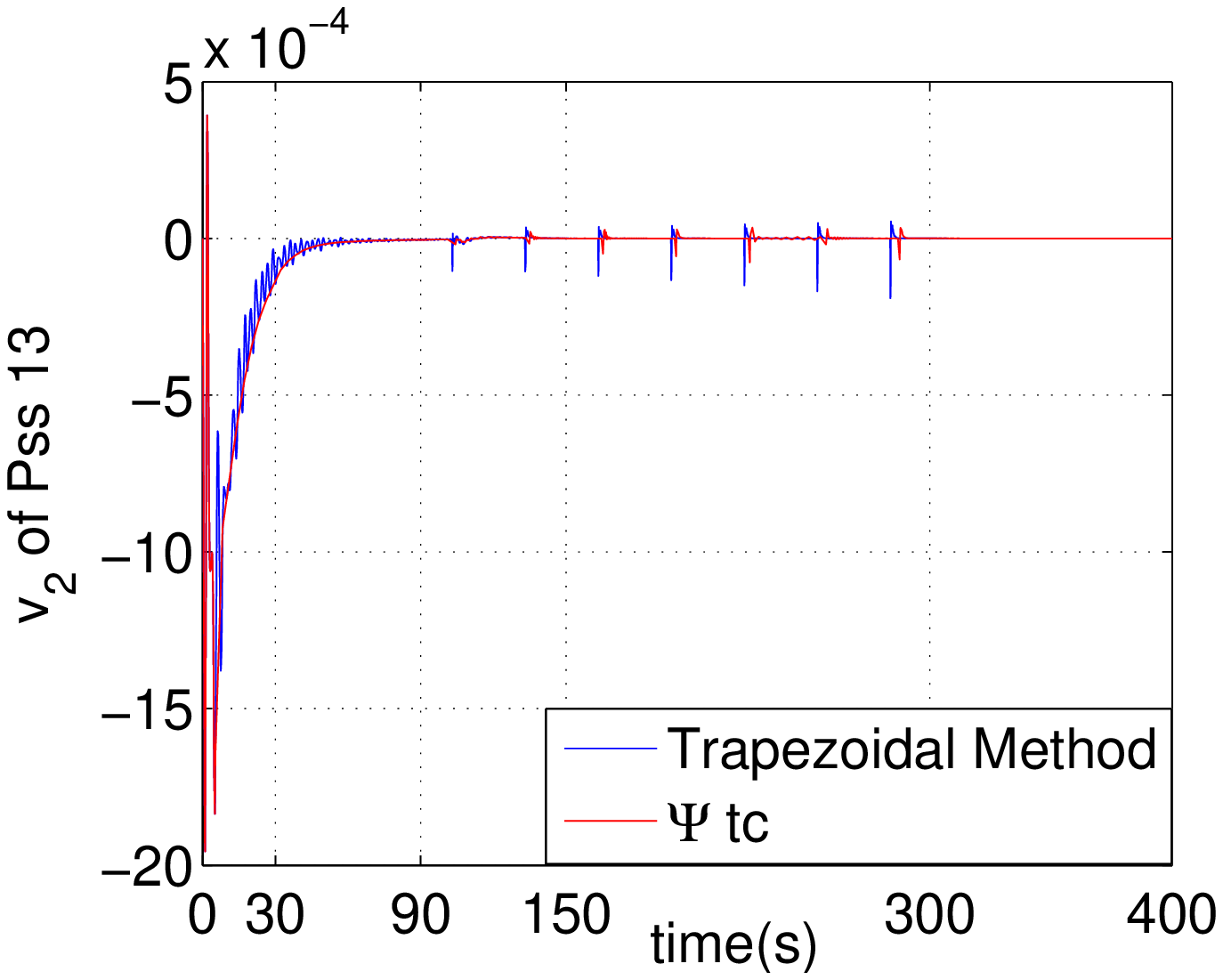}
\end{minipage}%
\begin{minipage}[t]{0.5\linewidth}
\includegraphics[width=1.8in ,keepaspectratio=true,angle=0]{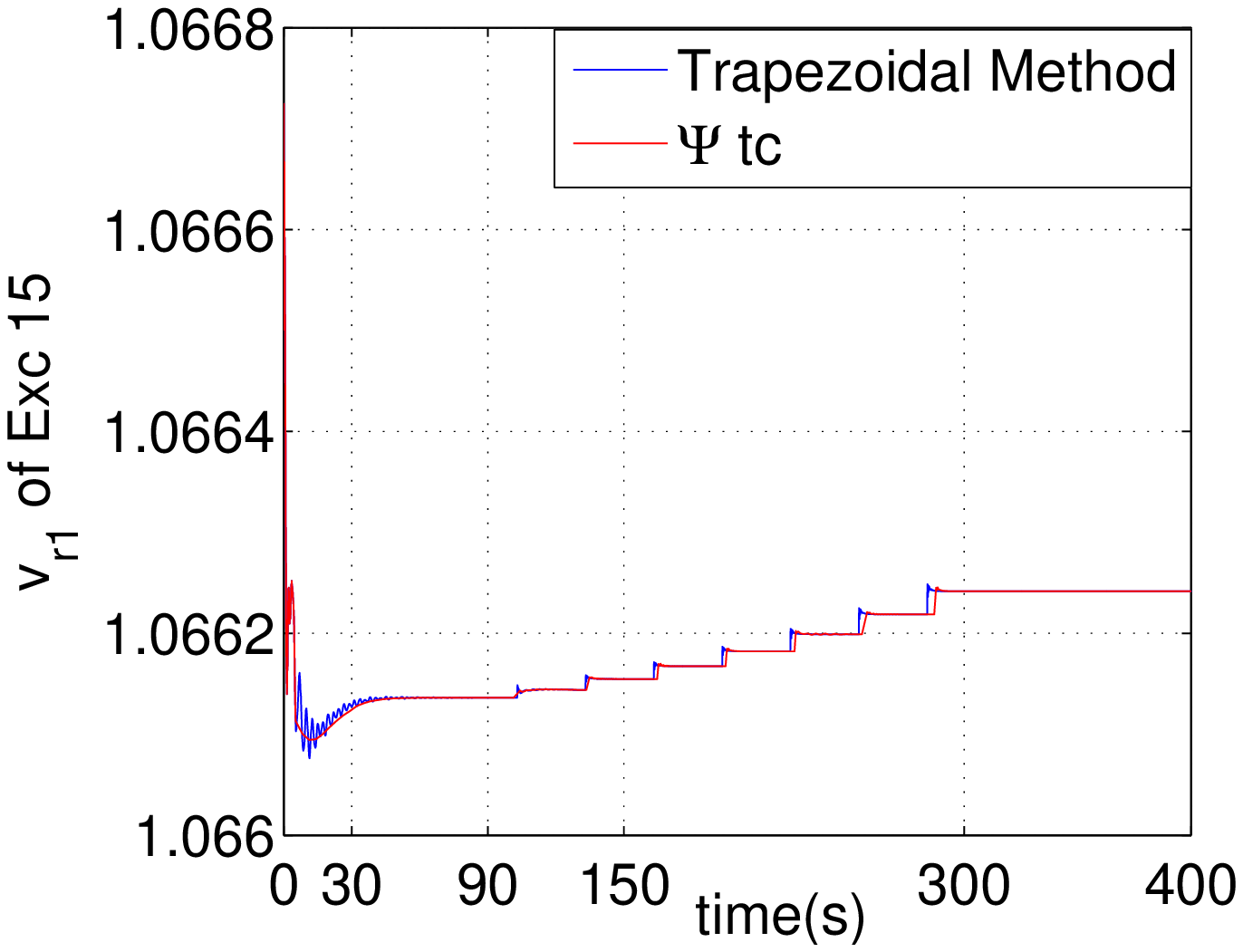}
\end{minipage}
\begin{minipage}[t]{0.5\linewidth}
\includegraphics[width=1.8in ,keepaspectratio=true,angle=0]{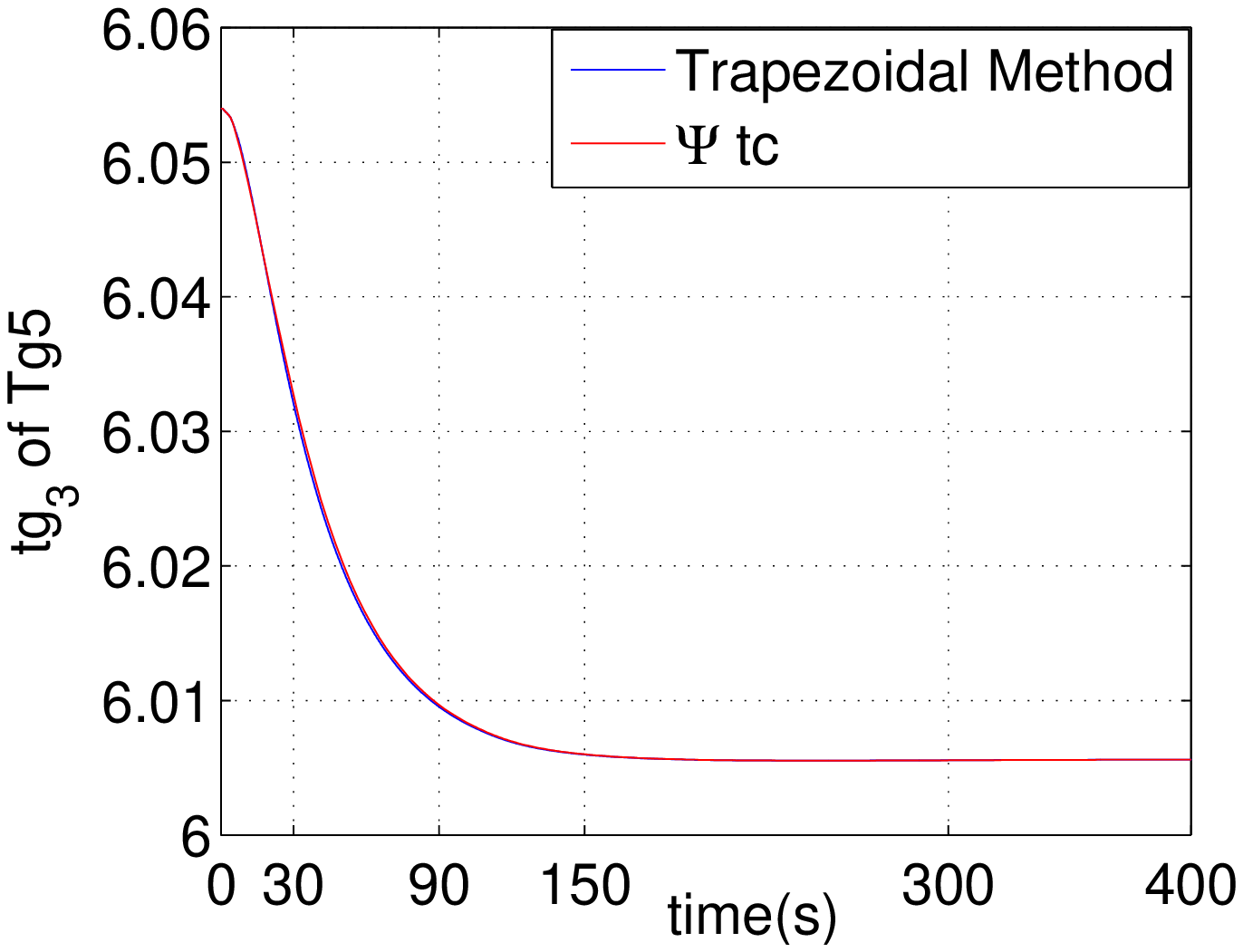}
\end{minipage}%
\begin{minipage}[t]{0.5\linewidth}
\includegraphics[width=1.8in ,keepaspectratio=true,angle=0]{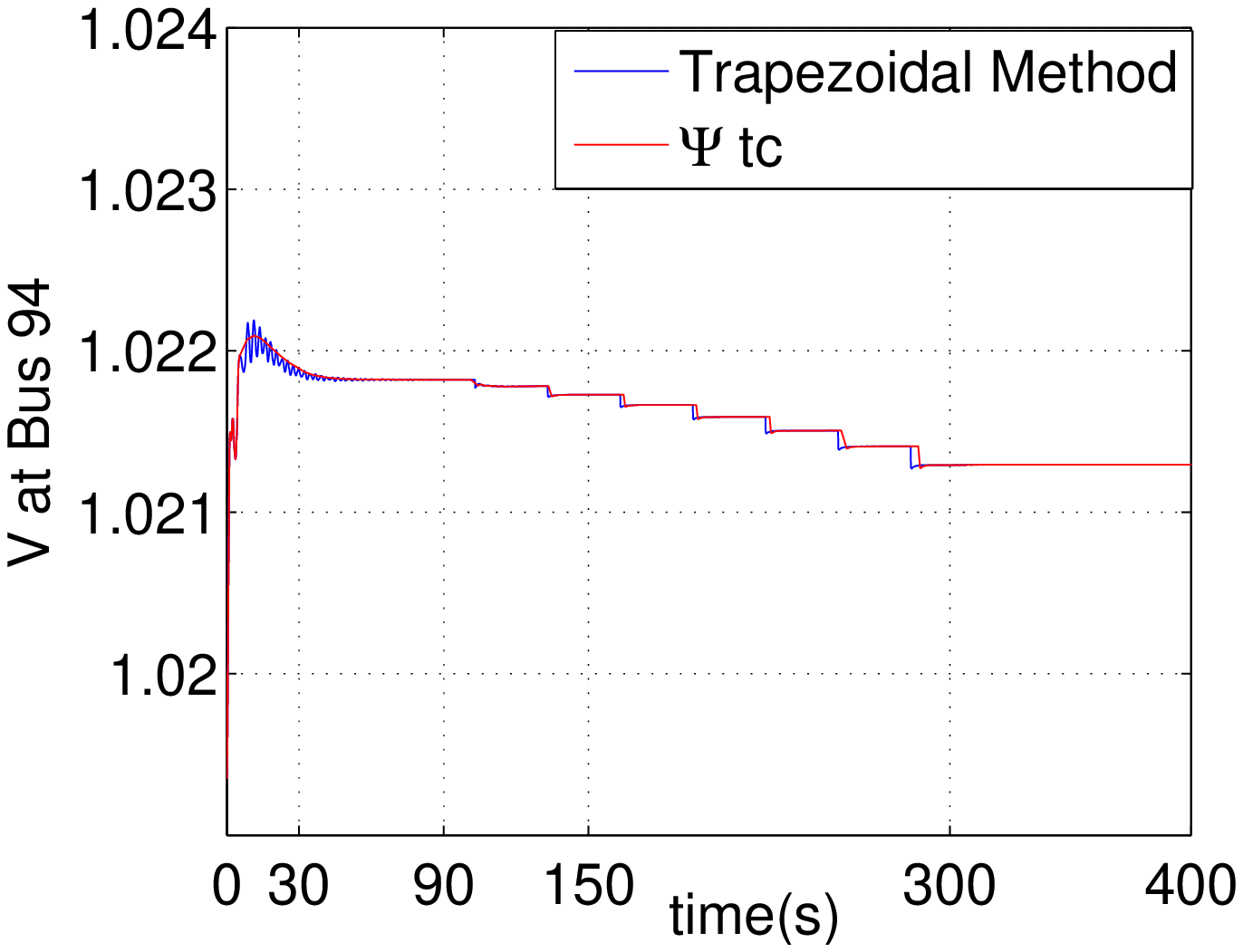}
\end{minipage}
\caption{The trajectory comparisons of the long-term stability model by implicit trapezoidal method and $\Psi tc$ method. $\Psi tc$ method provided correct approximations.}\label{TDSPTC}
\end{figure}

In this case, the post-fault system was stable after the contingency in the long-term time scale and it took 122.39s for the time domain simulation of the long-term stability model by implicit trapezoidal method. However, $\Psi tc$ method only took 16.12s for the simulation of the long-term stability model which was about $13.17\%$ of the time consumed by trapezoidal method. Fig. \ref{TDSPTC} shows that the trajectories by $\Psi tc$ method followed closely to the trajectories by trapezoidal method, and finally both converged to the same long-term stable equilibrium point. Thus $\Psi tc$ method provided correct approximations for the long-term stability model in terms of trajectories and stability assessment.

%Trapezoidal long-term stability model: 122.387910s
%PTC long-term stability model: 16.119617s.

\subsection{Numerical Example II}
\begin{figure}[!ht]
\centering
\begin{minipage}[t]{0.5\linewidth}
\includegraphics[width=1.8in ,keepaspectratio=true,angle=0]{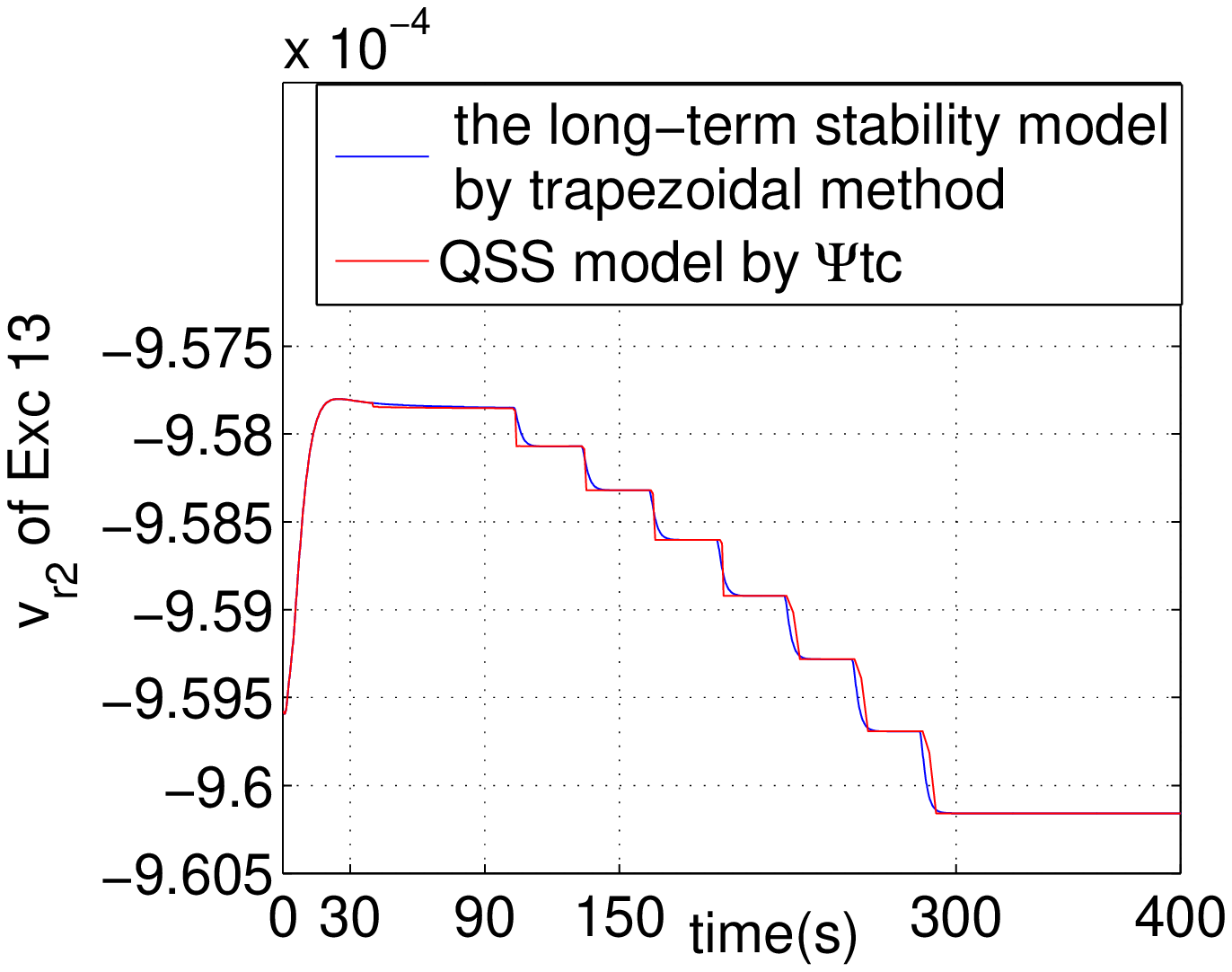}
\end{minipage}%
\begin{minipage}[t]{0.5\linewidth}
\includegraphics[width=1.8in ,keepaspectratio=true,angle=0]{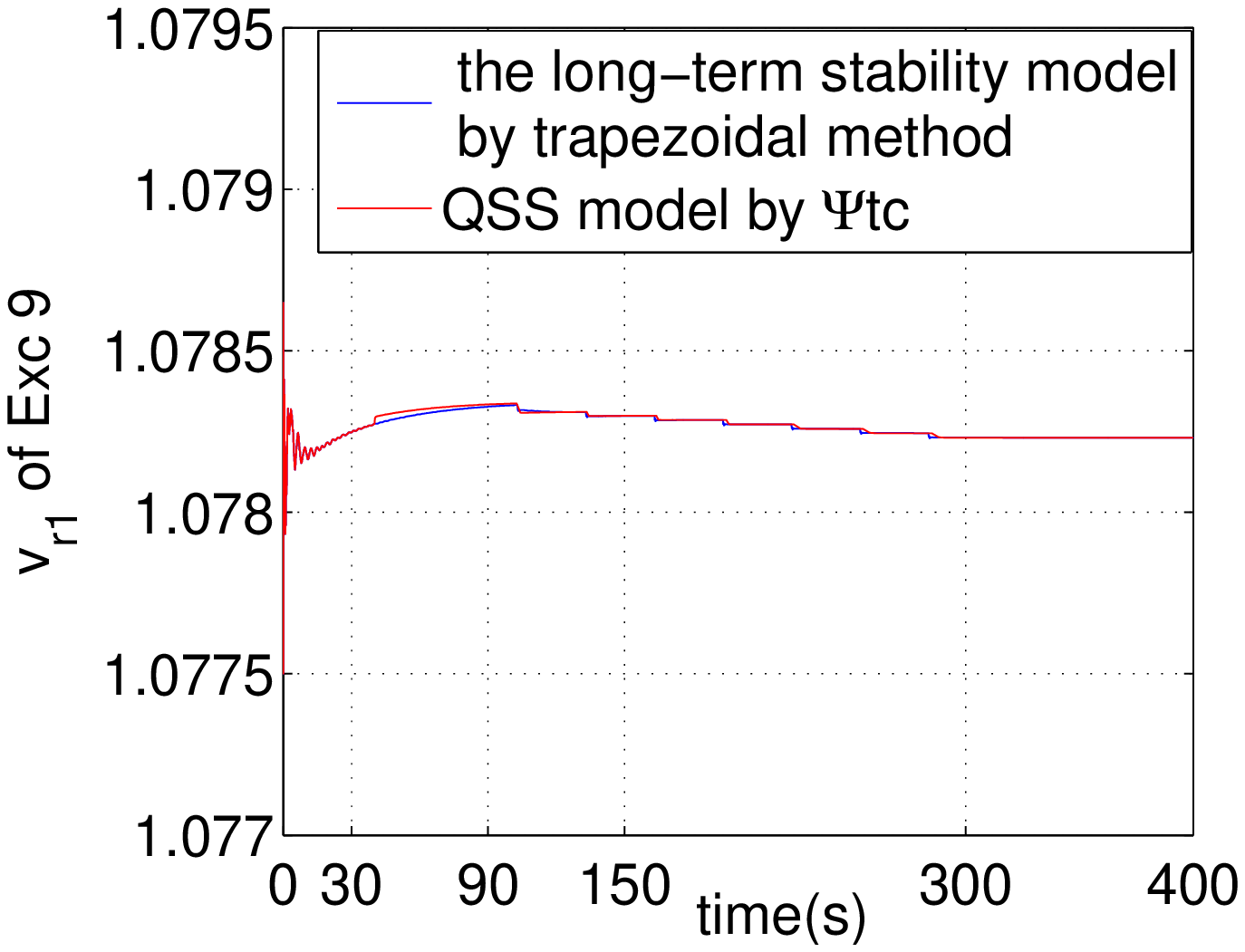}
\end{minipage}
\begin{minipage}[t]{0.5\linewidth}
\includegraphics[width=1.8in ,keepaspectratio=true,angle=0]{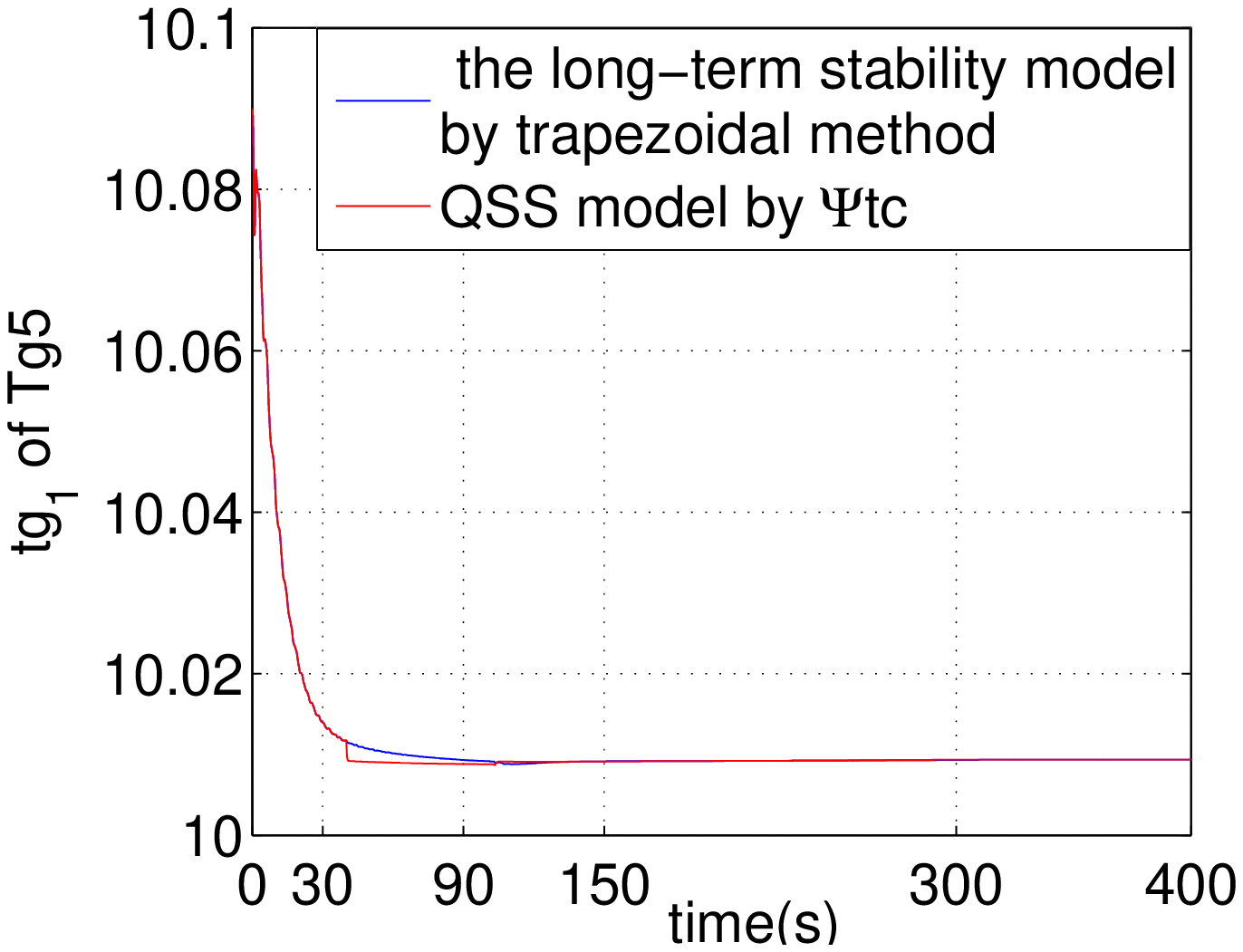}
\end{minipage}%
\begin{minipage}[t]{0.5\linewidth}
\includegraphics[width=1.8in ,keepaspectratio=true,angle=0]{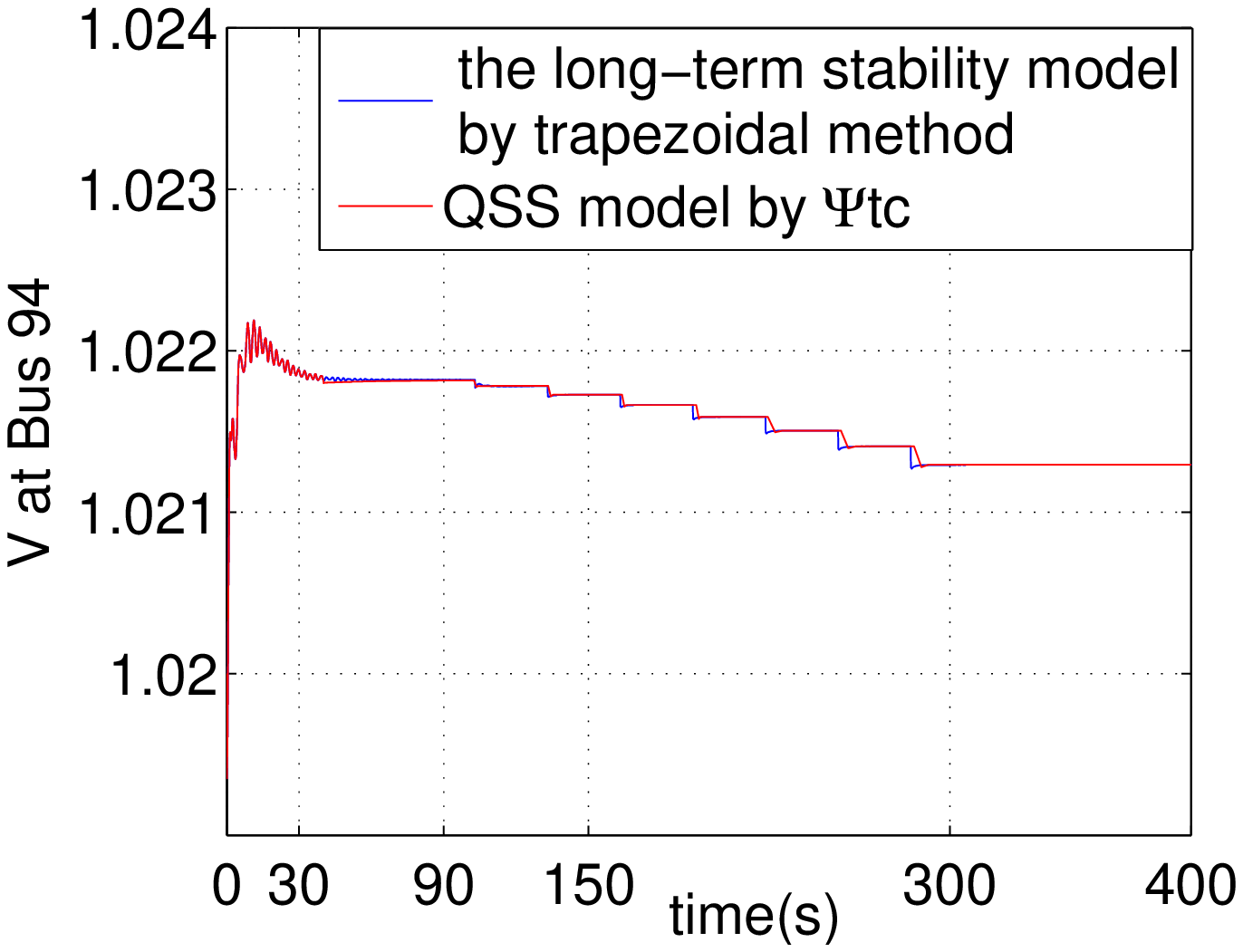}
\end{minipage}
\caption{The trajectory comparisons of the long-term stability model by implicit trapezoidal method and the QSS model by $\Psi tc$ method. $\Psi tc$ method overcame numerical difficulties and provided correct approximations.}\label{TDSPTCQSS}
\end{figure}

In this example, the system was the same as the last case. However, the QSS model met numerical difficulties at 40s when implicit trapezoidal method was used, thus $\Psi tc$ method was implemented in the QSS model starting from 40s. Fig. \ref{TDSPTCQSS} shows that the trajectories by $\Psi tc$ method converged to the long-term stable equilibrium point which the long-term stability model converged to, and also provided good accuracy for the intermittent trajectories. It took 21.75s for $\Psi tc$ method which  was about $17.77\%$ %is less than $\frac{1}{5}$
of the time consumed by the long-term stability model using implicit trapezoidal method.

%Trapezoidal long-term stability model: 122.387910s
%QSS fail and switch over to PTC: 21.752638s.

\subsection{Numerical Example III}
In this case, the 14-bus system was long-term unstable due to wild oscillations of fast variables. The system was modified based on the 14-bus test system in psat-2.1.6\cite{Milano:article}. There were three exponential recovery loads at Bus 9, 10 and 14 respectively and two turbine governors at Generator 1 and 3. Besides, there were over excitation limiters at all generators and three load tap changers at lines between Bus 4-9, Bus 12-13 and Bus 2-4. %The parameters of the system are given in the appendix.

The system suffered from long-term instabilities and simulation by implicit trapezoidal method could not continue after 101.22s. $\Psi tc$ method also stopped at 103.34s when the bound on the total number of iterations for the Newton step was reached. Thus $\Psi tc$ method was able to capture instabilities of the long-term stability model.
\begin{figure}[!ht]
\centering
\begin{minipage}[t]{0.5\linewidth}
\includegraphics[width=1.8in ,keepaspectratio=true,angle=0]{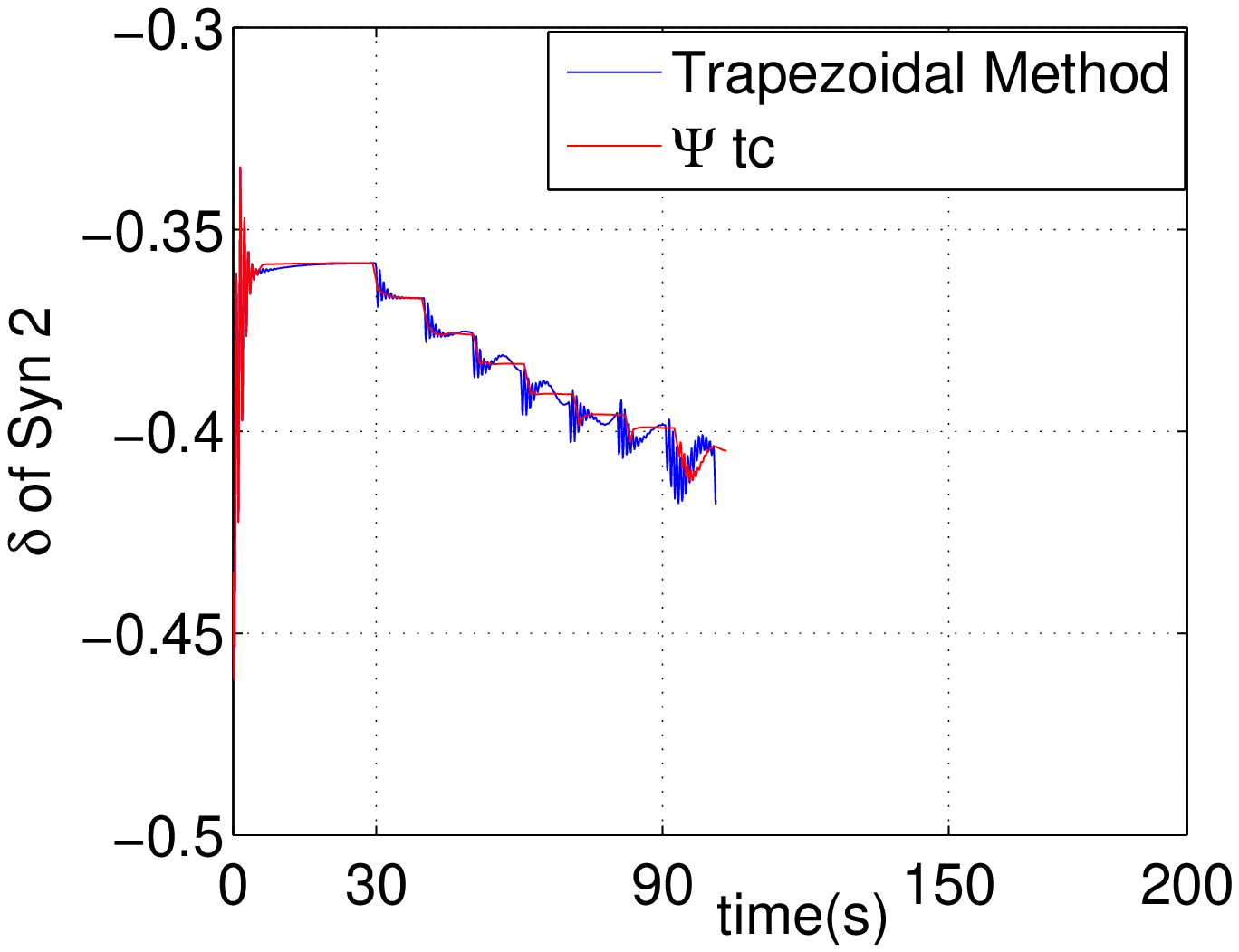}
\end{minipage}%
\begin{minipage}[t]{0.5\linewidth}
\includegraphics[width=1.8in ,keepaspectratio=true,angle=0]{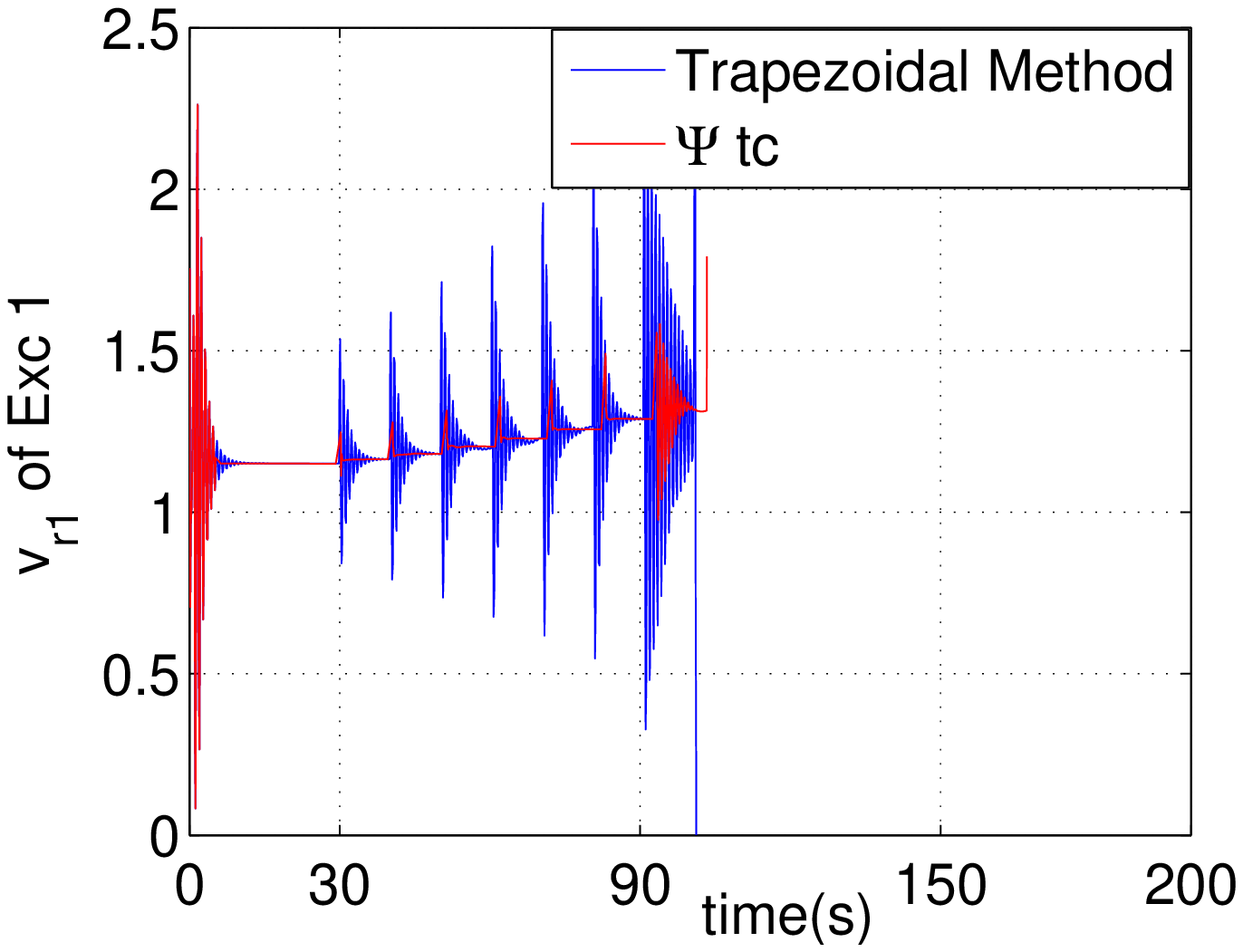}
\end{minipage}
\begin{minipage}[t]{0.5\linewidth}
\includegraphics[width=1.8in ,keepaspectratio=true,angle=0]{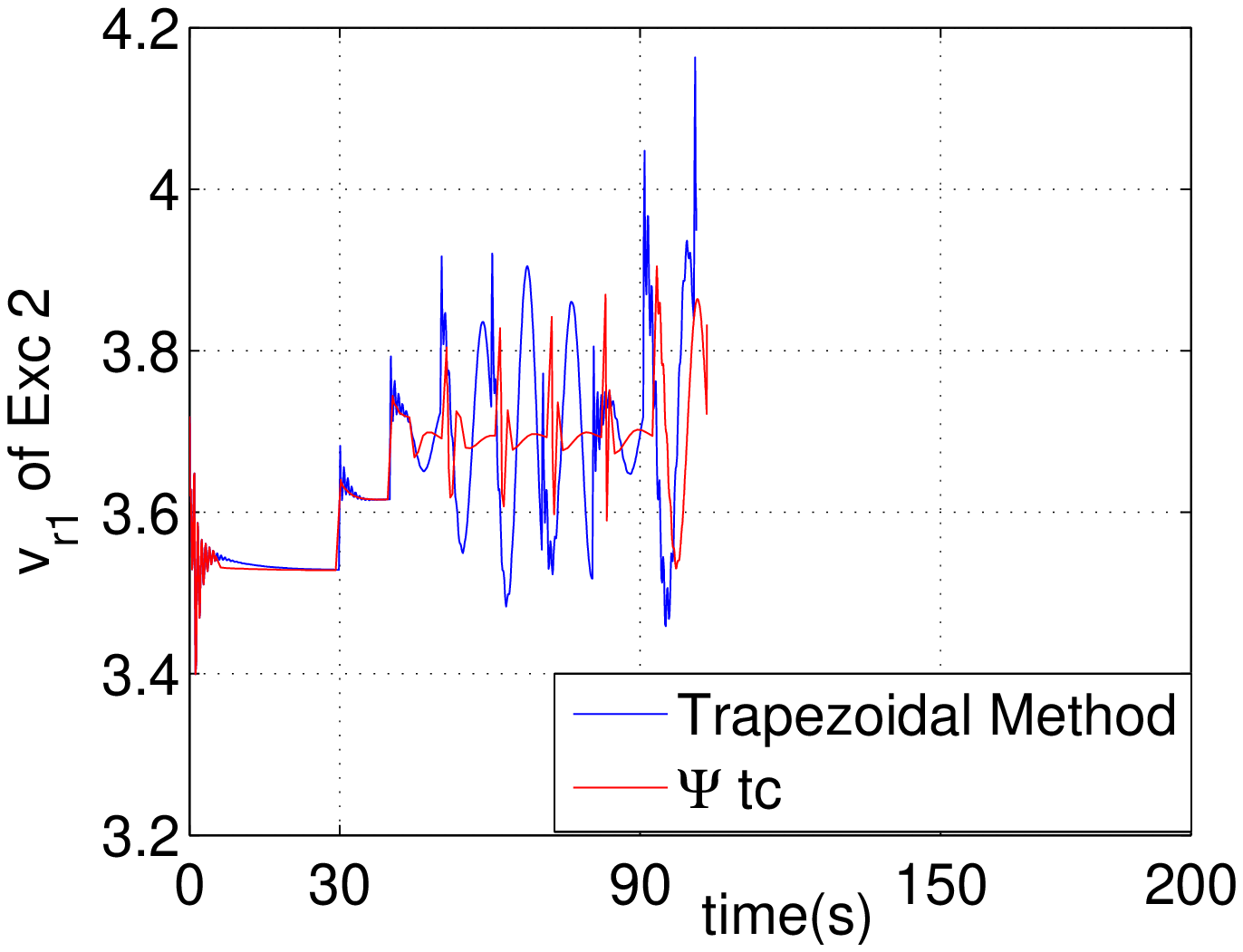}
\end{minipage}%
\begin{minipage}[t]{0.5\linewidth}
\includegraphics[width=1.8in ,keepaspectratio=true,angle=0]{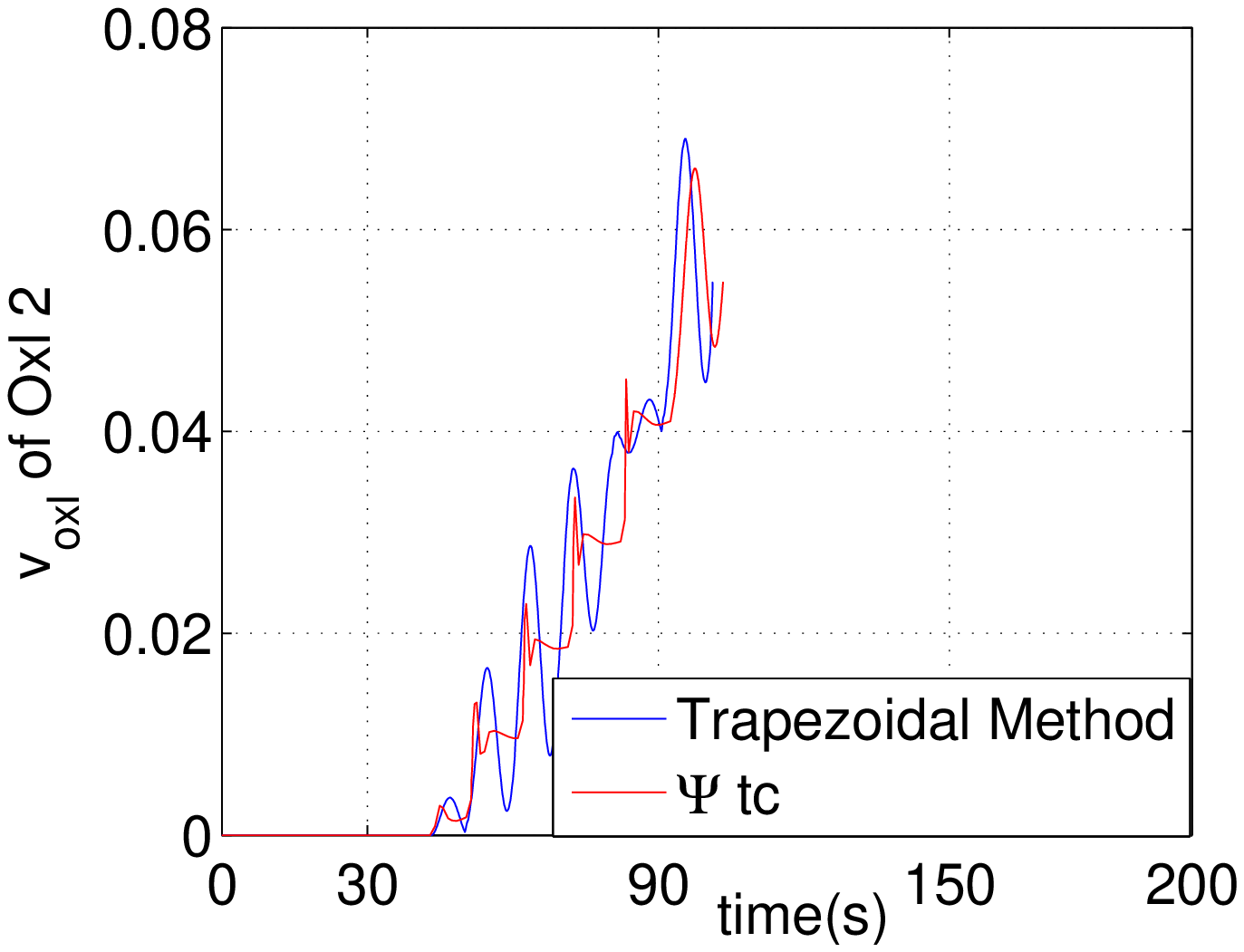}
\end{minipage}
\caption{The trajectory comparisons of the long-term stability model using implicit trapezoidal method and $\Psi tc$ method. $\Psi tc$ method was able to capture instabilities of the long-term stability model.}\label{my14completeqss_try}
\end{figure}

\section{Conclusion}\label{sectionconclusion}
In this paper, modified $\Psi tc$ methods for the long-term stability model and the QSS model are given for power system long-term stability analysis with illustrative numerical examples. We make use of the fast asymptotic convergence of $\Psi tc$ method in the long-term stability model to achieve fast simulation speed. On the other hand, we take advantage of good stability property of $\Psi tc$ method in the QSS model to overcome numerical difficulties. Numerical examples show that $\Psi tc$  can successfully provide correct stability assessment for the long-term stability model and overcome numerical difficulties in the QSS model, as well as offer good accuracy for the intermediate trajectories. $\Psi tc$ can be regarded as a good in-between method with respect to integration and steady state calculation, thus serves as an alternative method in power system long-term stability analysis.

\section{Acknowledgment}
This work was supported by the Consortium for Electric Reliability Technology Solutions provided by U.S. Department No. DE-FC26-09NT43321.

\appendices
\ifCLASSOPTIONcaptionsoff
  \newpage
\fi

\end{document}